\documentclass[prd,floats,floatfix,nofootinbib]{revtex4}
\usepackage[dvips]{graphicx}
\usepackage{graphics}
\usepackage{color}

\begin{document}

\title{Buchdahl Limit and TOV Equations in Interacting Vacuum Scenarios}

\author{Rodrigo Maier\footnote{rodrigo.maier@uerj.br}} 

\affiliation{
Departamento de F\'isica Te\'orica, Instituto de F\'isica, Universidade do Estado do Rio de Janeiro,\\
Rua S\~ao Francisco Xavier 524, Maracan\~a,\\
CEP20550-900, Rio de Janeiro, Brasil
}

\date{\today}

\begin{abstract}
We investigate the stability of ultra-compact stellar configurations in the context of an interacting vacuum component. By extending the Tolman-Oppenheimer-Volkoff equations to include a covariant energy exchange between the fluid and vacuum sectors, we examine how the classical Buchdahl stability limit is modified. We analyze two phenomenological interaction models: a coupling to the matter energy density gradient and a direct coupling to the spacetime curvature. Numerical integration reveals that while standard General Relativity predicts a central pressure divergence as the compactness approaches the Buchdahl threshold, the interaction term $Q_\nu$ relaxes the pressure gradient and maintains a finite, well-behaved central pressure for proper domains of the coupling parameter. These results demonstrate that an interacting vacuum provides a physical mechanism to bypass classical geometric bounds, potentially supporting ultra-compact objects in regimes previously considered singular.
\end{abstract}

\maketitle

\section{Introduction}

The study of stellar equilibrium in General Relativity has remained a cornerstone of relativistic astrophysics since the formalization of the theory. Central to this field are the Tolman-Oppenheimer-Volkoff (TOV) equations, which dictate the hydrostatic equilibrium of a static, spherically symmetric fluid body\cite{Tolman:1939jz,Oppenheimer:1939ne}. A profound consequence of these equations is the existence of the Buchdahl limit\cite{Buchdahl:1959zz} which establishes that for a Schwarzschild interior solution with a non-increasing density profile, the mass-to-radius ratio must satisfy $M/R \leq 4/9$. This limit represents a fundamental stability threshold beyond which no internal pressure can prevent gravitational collapse. In the modern era, the state of the art in this field has been significantly reshaped by high-precision observations from the Laser Interferometer Gravitational-Wave Observatory (LIGO) and the Neutron Star Interior Composition Explorer (NICER). 
Through the detection of binary neutron star mergers\cite{LIGOScientific:2018cki} and the precise mapping of pulsar mass-radius relations for both canonical and high-mass sources\cite{Miller:2019cac,Miller:2021qha}, these missions have provided unprecedented constraints on the neutron star equation of state (EoS), necessitating a rigorous re-evaluation of classical stability limits in the high-density regime.

Despite the remarkable success of General Relativity, the nature of dark energy and the mechanism driving late-time cosmic acceleration remain among the most significant challenges in modern physics. Specifically, the cosmological constant problem -- the immense discrepancy between the observed vacuum energy density and quantum field theory predictions\cite{Weinberg:1988cp} -- and the cosmic coincidence problem\cite{Zlatev:1998tr}, which questions why the energy densities of matter and dark energy are of the same order today, present deep theoretical tensions. These puzzles, compounded by the persistent $H_0$ tension between early- and late-universe measurements\cite{DiValentino:2021izs}, have motivated the study of modified gravity and alternative energy-momentum sectors. In this context, frameworks such as $f(R)$ gravity, scalar-tensor theories, and Gauss-Bonnet gravity introduce additional geometric degrees of freedom that modify the TOV equations and, consequently, the Buchdahl limit\cite{Astashenok:2013vza}. Research has shown that in modified gravity the $4/9$ threshold can be shifted or even relaxed\cite{Capozziello:2011nr}, potentially allowing for ultra-compact objects that would be unstable under standard General Relativity. Furthermore, the generalization of these limits now frequently accounts for anisotropic pressures ($P_r \neq P_t$) and electromagnetic charges, both of which play a vital role in the stability of exotic compact objects\cite{Mak:2001eb,Guilfoyle:1999yb}.

Parallel to these geometric modifications is the increasingly popular interacting vacuum scenario. Unlike the rigid cosmological constant $\Lambda$, this model proposes a dynamical vacuum energy density that exchanges energy and momentum with the matter sector. This approach is motivated by its potential to resolve the coincidence problem and alleviate the $H_0$ tension by allowing for a vacuum that evolves alongside the universe's large-scale structure\cite{Wang:2016lxa,Salvatelli:2014zta}. In the context of compact objects, this interaction implies that the stress-energy tensor is no longer independently conserved, $\nabla_\mu T^{\mu\nu} \neq 0$, but is instead coupled to an interaction term $Q^\nu$. Recent studies in running vacuum models and energy-momentum-squared gravity have begun to explore how such couplings alter the internal structure of stars, potentially preventing the formation of singularities\cite{SolaPeracaula:2021gxi,Board:2017ign}.

In this paper, we bridge these two frontiers by examining the Buchdahl limit and the TOV equations within the context of an interacting vacuum component. By deriving the modified equilibrium conditions, we investigate how the energy exchange between the fluid and the vacuum influences the maximum compactness and stability of static configurations. This analysis aims to provide new insights into the constraints on vacuum interactions and their implications for the existence of compact objects in the strong-field regime.

We organize this paper as follows.
In the next section we extend the TOV equation for an arbitrary interacting vacuum component. 
Assuming that the average vacuum density exhibits a monotonic decrease toward the stellar surface
we also show that standard Buchdahl inequality with a cosmological constant is restored. 
In section III we provide two neat covariant choices for the transfer vector $Q^\nu$. For proper domains of the parameters we show explicit violations of the classical Buchdahl limit with a nonvanishing cosmological constant.
In section IV we assume a proper choice for $Q^\mu$ from pure phenomenological grounds in order to analyze how the 
modified pressure gradients and resulting interior solutions shift the fundamental $M/R \leq 4/9$ stability
threshold.
In section V we leave our final remarks.

\section{Extended TOV Equation and Buchdahl Inequalities}
We start by considering the field equations
\begin{eqnarray}
\label{eq2}
G_{\mu\nu}= 8\pi (T_{\mu\nu}-V g_{\mu\nu})
\end{eqnarray}
where $T_{\mu\nu}$ is the energy-momentum tensor which comes from a perfect fluid 
\begin{eqnarray}
\label{eq3}
T^{\mu}_{~~\nu}&=&(\rho+p)u^\mu u_\nu+ p \delta^{\mu}_{~\nu},
\end{eqnarray}
which interacts with the vacuum component $V$ through a $4$-vector $Q_\nu$,
namely
\begin{eqnarray}
\label{eq4}
\nabla_\mu (T^\mu_{~\nu})=\nabla_\nu V=Q_\nu.    
\end{eqnarray}
so that Bianchi identities are automatically satisfied.

Let us now assume a general static spherically symmetric geometry in the coordinates $(t, r, \theta, \phi)$ with line element
\begin{eqnarray}
ds^2=-f(r)dt^2+h(r)dr^2+r^2(d\theta^2+\sin^2\theta d\phi^2).    
\end{eqnarray}
Fixing $u^\mu=-(1/\sqrt{f})\delta^\mu_{~t}$, the field equations (\ref{eq2}) together with the Bianchi equations (\ref{eq4}) furnish
\begin{eqnarray}
\label{eq00}
&&\frac{1}{r^2}\frac{d}{dr}\Big[r\Big(1-\frac{1}{h}\Big)\Big]=8\pi(\rho+V),\\
\label{eq11}
&&-\frac{1}{r^2}\Big(1-\frac{1}{h}\Big)+\frac{f^\prime}{rfh}=8\pi(p-V),\\
\label{eq33}
&&\frac{1}{2}\Big[\frac{1}{\sqrt{fh}}\frac{d}{dr}\Big(\frac{f^\prime}{\sqrt{fh}}\Big)+\frac{f^\prime}{rfh}-\frac{h^\prime}{rh^2}\Big]=8\pi(p-V),\\
\label{eqcon}
&&p^\prime=-\frac{f^\prime}{2f}(\rho+p)+V^\prime.
\end{eqnarray}
At this stage we introduce standard definition
of mass 
%
\begin{eqnarray}
M(r)=4\pi\int^r\rho(\xi)\xi^2d\xi.
\end{eqnarray}
Inspired by the above it turns to be useful to include the definition
\begin{eqnarray}
{\cal V}=4\pi\int^rV(\xi)\xi^2d\xi.    
\end{eqnarray}
From (\ref{eq00}) we then obtain
\begin{eqnarray}
\label{hsol}
h(r)=\Big\{1-2\Big[\frac{M(r)+{\cal V}(r)}{r}\Big]\Big\}^{-1}.   
\end{eqnarray}
An extension of the Tolman-Oppenheimer-Volkoff equation can now be derived once one takes into account the above solution together with a direct substitution of (\ref{eq11}) in (\ref{eqcon}):
\begin{eqnarray}
\label{tov}
\frac{dp}{dr}=-(\rho+p)\Big\{\frac{M(r)+{\cal V}(r)+4\pi r^3(p-V)}{r[r-2(M(r)+{\cal V}(r))]}\Big\}+{V^\prime}.~~~~    
\end{eqnarray}
It is easy to see that (\ref{tov}) reduces to the TOV-$\Lambda$ equation\cite{Boehmer:2003uz} for a constant vacuum component.

In order to examine the Buchdahl inequality in the present context one may consider the
subtraction of (\ref{eq33}) from (\ref{eq11})  which furnishes
\begin{eqnarray}
\label{w1n}
\nonumber
\frac{d}{dr}\Big[\frac{1}{r\sqrt{h}}\frac{d}{dr}(\sqrt{f})\Big]=\sqrt{fh}
\Big[\frac{1}{r^3}\Big(\frac{1}{h}-1\Big)+\frac{1}{2r^2h^2}\Big(\frac{dh}{dr}\Big)\Big].
\end{eqnarray}
Substituting (\ref{hsol}) in the brackets of the RHS above, we obtain
\begin{eqnarray}
\label{w1}
\frac{d}{dr}\Big[\frac{1}{r\sqrt{h}}\frac{d}{dr}(\sqrt{f})\Big]=\sqrt{fh}\frac{d}{dr}(\bar{\rho}+\bar{v}),   
\end{eqnarray}
where we have defined the average energy and vacuum densities as
\begin{eqnarray}
\bar{\rho}=\frac{M(r)}{r^3},~~\bar{v}=\frac{{\cal V}(r)}{r^3},    
\end{eqnarray}
respectively. At this stage two basic assumptions may be considered: 
\\
\\
(i) As in the original noninteracting case we shall assume that 
the radial energy density rate decreases, namely $d\rho/dr\leq0$. Therefore, the average density $\bar{\rho}$ must also decrease monotonically as $r$ increases; 
\\
\\
(ii) In the same vein of the above hypothesis one might assume that the average vacuum density follows a similar behaviour, namely $dV/dr \leq 0$, so that ${\bar v}$ also decreases monotonically with $r$.
\\
\\
The second premise regarding the behaviour of the average vacuum density seems to be 
reasonable as long as one assumes that the vacuum component is positive definite. In fact, 
let $\gamma$ denote the boundary of the matter distribution. In this case it is worth to note that $V(\gamma)$ account for a cosmological constant, say $V_0$, so that the interior solution can be matched with the known Schwarschild-de Sitter spacetime in which for $r\geq \gamma$
\begin{eqnarray}
f(r)=\frac{1}{h(r)}=1-\frac{2M}{r}-\frac{8\pi V_0 r^2}{3}.    
\end{eqnarray}
As $V_0$ is connected to the usual cosmological constant one might assume that, at the boundary, $V_0$ accounts for a negligible contribution.
As we move towards the center of the matter distribution the vacuum component $V$ is expected to interact with the matter sector given the coupling in equation (\ref{eqcon}). In this case the contribution of $V(r)$ may be enhanced so that its radial rate decreases as we move towards the boundary. As $dV/dr\leq 0$, its average density $\bar {v}$ must also decrease monotonically as $r$ increases. In the case that both assumptions above are satisfied we end up with a Buchdalh inequality analogous to the case of a constant vacuum component. 
In fact, from (\ref{w1}) we obtain
\begin{eqnarray}
\label{in1}
\frac{d}{dr}\Big[\frac{1}{r\sqrt{h}}\frac{d}{dr}(\sqrt{f})\Big]\leq0.  
\end{eqnarray}
A direct inward integration of (\ref{in1}) from the surface $r=\gamma$ gives us
\begin{eqnarray}
\label{eq17}
\frac{1}{r\sqrt{h}}\frac{d}{dr}(\sqrt{f})\geq  \frac{M}{\gamma^3}-\frac{8\pi V_0}{3}.  
\end{eqnarray}
A further integration of the above then furnishes
\begin{eqnarray}
\nonumber
\sqrt{f(0)}\leq \Big(1-\frac{2M}{\gamma}-\frac{8\pi V_0 \gamma^2}{3}\Big)^{\frac{1}{2}}-\Big(\frac{M}{\gamma^3}-\frac{8\pi V_0}{3}\Big)\\
\label{in3}
\times\int_0^\gamma r
\Big\{1-2\Big[\frac{M(r)+{\cal V}(r)}{r}\Big]\Big\}^{-\frac{1}{2}}dr.
\end{eqnarray}

Finally, one may note that (i) and (ii) imply that
%
\begin{eqnarray}
\label{in4n}
\bar{\rho}\geq \frac{M}{\gamma^3}~~{\rm and}~~{\bar v}\geq \frac{4\pi V_0}{3},    
\end{eqnarray}
%
so that 
the second term in the RHS of (\ref{in3}) will acquire its smallest value once the equalities in (\ref{in4n}) hold.
In this case we obtain
\begin{eqnarray}
\nonumber
\sqrt{f(0)}\leq \Big(1-\frac{2M}{\gamma}-\frac{8\pi V_0 \gamma^2}{3}\Big)^{\frac{1}{2}}-\Big(\frac{M}{\gamma^3}-\frac{8\pi V_0}{3}\Big)\\
\label{in3q}
\times\int_0^\gamma r
\Big[1-\frac{2Mr^2}{\gamma^3}-\frac{8\pi V_0 r^2}{3}\Big]^{-\frac{1}{2}}dr.
\end{eqnarray}
As $\sqrt{f(0)}\geq 0$, the above condition furnishes
\begin{eqnarray}
\label{bds}
\Big(1-\frac{2M}{\gamma}-\frac{8\pi V_0 \gamma^2}{3}\Big)^{\frac{1}{2}} \geq    \frac{1}{3}-\frac{8\pi V_0 \gamma^3}{9M},
\end{eqnarray}
as one should expect\cite{Boehmer:2002gg,Boehmer:2003uz,Boehmer:2025uoq}.

\section{Neat Covariant Violations of Buchdahl Limit}
In this section, we present two distinct examples illustrating how an
interacting vacuum component can lead to violations of the classical 
Buchdahl inequality (\ref{bds}). To explore this, we consider two specific 
covariant formulations for the interaction 4-vector $Q_\nu$:
\begin{eqnarray}
^{(1)}Q_\nu &=& \chi \nabla_\nu(T_{\alpha\beta}u^\alpha u^\beta),\\ 
^{(2)}Q_\nu &=& \chi \nabla_\nu R.
\end{eqnarray}
By exploring these two specific cases, we aim to provide a 
phenomenological survey of how different coupling mechanisms influence the 
stellar equilibrium. 

The first choice, $^{(1)}Q_\nu$, represents a direct interaction with the
material source, where the energy exchange is driven by gradients in the 
local energy density. This approach allows us to track how the vacuum 
reacts to the presence of matter concentrations, potentially acting as a 
local regulatory mechanism.

In contrast, the second choice, $^{(2)}Q_\nu$, couples the interaction 
directly to the spacetime curvature via the Ricci scalar. This is 
particularly compelling as it ensures the vacuum interaction is 
intrinsically linked to the gravitational field itself, effectively 
switching off in the limit of flat spacetime where $R=0$.

Beyond the fundamental requirement of general covariance, the selection of
the interaction 4-vectors $^{(1)}Q_\nu$ and $^{(2)}Q_\nu$ is grounded in 
physical frameworks that treat the vacuum as a dynamical participant in 
the universe's evolution. Specifically, the choice $^{(1)}Q_\nu \propto 
\nabla_\nu \rho$ (where $\rho = T_{\alpha\beta}u^\alpha u^\beta$) models 
the vacuum energy density as a functional of the local matter density. 
Similar density-dependent couplings are prevalent in phenomenological 
literature where the static cosmological constant $\Lambda$ is replaced by 
a dynamical component that exchanges energy with a cold dark matter fluid 
to ensure the conservation of the total stress-energy tensor. Notably, 
such interactions are often motivated by their ability to furnish 
nonsingular cosmological configurations, offering a theoretical pathway to 
avoid the initial Big Bang singularity \cite{Bruni:2021msx,Alves:2024pyz}. 
Furthermore, these models have been shown to support structure formation 
within an accelerated cosmic background by allowing perturbations to grow 
over time \cite{Munyeshyaka:2022fck}, with recent cosmological constraints 
discussed in \cite{Yang:2025boq}. 
Conversely, the interaction $^{(2)}Q_\nu 
\propto \nabla_\nu R$ couples the vacuum directly to the Ricci scalar. 
This formulation is closely aligned with the predictions of semi-classical 
gravity and Running Vacuum Models (RVM), where the vacuum energy density 
$\rho_{\text{vac}}$ is expected to evolve as a function of the energy 
scale. Within an FLRW framework, this dependence is typically expressed 
through the Hubble rate and its time derivative, $H^2$ and $\dot{H}$ (see 
\cite{SolaPeracaula:2023swx} and references therein).

\subsection{$Q_\nu={^{(1)}}Q_\nu$}
We begin our analysis by considering the first interaction model, defined by the coupling to the gradient of the matter energy density.
In this case a direct integration of (\ref{eq4}) furnishes
\begin{eqnarray}
\label{solv}
V(r)=V_0+\chi \rho(r).    
\end{eqnarray}
so that
\begin{eqnarray}
\label{w1mm}
{\cal V}(r)=\frac{4}{3}\pi V_0 r^3+\chi M(r) \rightarrow \bar{v}= \frac{4}{3}\pi V_0+\chi \bar{\rho}.    
\end{eqnarray}
In this case the metric of the exterior Schwarzschild de-Sitter spacetime exhibits a mass shift in the metric, namely, for $r \geq \gamma$
\begin{eqnarray}
f(r)=\frac{1}{h(r)}=1-\frac{2M(1+\chi)}{r}-\frac{8\pi V_0 r^2}{3}.    
\end{eqnarray}

From Einstein field equations one may show that (\ref{w1}) reduces to
\begin{eqnarray}
\label{w1m}
\frac{d}{dr}\Big[\frac{1}{r\sqrt{h}}\frac{d}{dr}(\sqrt{f})\Big]=\sqrt{fh}(1+\chi)\frac{d\bar{\rho}}{dr},   
\end{eqnarray}    
so that the classical Buchdahl limit (\ref{bds}) can be recovered in the case that
\begin{eqnarray}
\label{w1mp}
(1+\chi)\frac{d\bar{\rho}}{dr}\leq 0.    
\end{eqnarray}
Considering the case in which the average energy density rate decreases monotonically toward the stellar surface
a compactness limit analogous to that of the classical Buchdahl limit should be obtained as
long as $\chi\geq -1$ so that (\ref{in1}) still holds.
In this case, one may show that (\ref{eq17}) turns into
\begin{eqnarray}
\label{eq17n}
\frac{1}{r\sqrt{h}}\frac{d}{dr}(\sqrt{f})\geq  \frac{M(1+\chi)}{\gamma^3}-\frac{8\pi V_0}{3}.  
\end{eqnarray}
A further integration of the above then furnishes
\begin{eqnarray}
\nonumber
\sqrt{f(0)}\leq \Big[1-\frac{2M(1+\chi)}{\gamma}-\frac{8\pi V_0 \gamma^2}{3}\Big]^{\frac{1}{2}}~~~~~~~~~~~~~~~~~~\\
\label{in3}
-\Big[\frac{M(1+\chi)}{\gamma^3}-\frac{8\pi V_0}{3}\Big]~~~~~~~~~~~~~~~~~~~~~~~~~~~~~~\\
\nonumber
\times\int_0^\gamma r
\Big\{1-2\Big[\frac{M(r)+{\cal V}(r)}{r}\Big]\Big\}^{-\frac{1}{2}}dr.~~~~~~
\end{eqnarray}
so that (\ref{in3q}) extends to
\begin{eqnarray}
\nonumber
\sqrt{f(0)}\leq \Big[1-\frac{2M(1+\chi)}{\gamma}-\frac{8\pi V_0 \gamma^2}{3}\Big]^{\frac{1}{2}}~~~~~~~~~~~~~~\\
\label{in3}
-\Big[\frac{M(1+\chi)}{\gamma^3}-\frac{8\pi V_0}{3}\Big]~~~~~~~~~~~~~~~~~~~~~~~~~~\\
\nonumber
~~~~~~~~~~~~\times\int_0^\gamma r
\Big\{1-\frac{2(1+\chi)M r^2}{\gamma^3}-\frac{8\pi V_0 r^2}{3}\Big\}^{-\frac{1}{2}}dr.
\end{eqnarray}
Again, %
as $\sqrt{f(0)}\geq 0$, the above condition furnishes the modified limit
\begin{eqnarray}
\label{mbl}
\sqrt{1-\frac{2M(1+\chi)}{\gamma}-\frac{8\pi V_0 \gamma^2}{3}}\geq
\frac{1}{3}-\frac{8\pi V_0 \gamma^3}{9M(1+\chi)}.
\end{eqnarray}
%
%

In the following, we shall give a numerical example of a violation of the Buchdahl bound by considering the idealized case of constant energy density $\rho\equiv \rho_0$, thereby isolating the influence of the vacuum interaction from the effects of internal density gradients.
Therefore, substituting (\ref{solv}) in the general TOV equation (\ref{tov}) we end up with
\begin{eqnarray}
\label{tov1}
\frac{dp}{dr}=-4\pi r(\rho_0+p)\Big\{\frac{(1-2\chi)\rho_0 + 3p-2V_0}{3-8\pi r^2 [V_0+
\rho_0(1+\chi)]}\Big\}.    
\end{eqnarray}

To test for violations of the Buchdahl limit, we shall perform a numerical inward integration of the internal pressure profile, starting from the vacuum boundary at the stellar surface. In the classical regime, the pressure diverges at the center as the compactness approaches the equality threshold of (\ref{bds}). However, if our modified field equations yield a finite central pressure for configurations exceeding this compactness, it serves as a definitive signature that the interaction term $Q_\nu$ has successfully bypassed the classical Buchdahl inequality.
In order to provide a clear benchmark for our results, we first specify the fundamental parameters of the model. To this end we fix the constant energy density $\rho_0=0.10$ and vary the vacuum energy density $V_0$ to generate a sequence of internal pressure profiles until the pressure approaches the classical limit. For the purposes of this numerical analysis, we shall assume stellar configurations of unit radius, that is $\gamma = 1$. This normalization allows us to precisely quantify how the introduction of a non-zero coupling $\chi$ shifts the stability threshold for a given mass, directly revealing the departures from the classical Buchdahl bound. In Fig. 1 we show the internal pressure profile for different values of $V_0$. Here we see that when $V_0\simeq 0.012$ the internal pressure diverges indicating that the classical Buchdahl limit has been reached.
\begin{figure}[tbp]
\includegraphics[width=8cm,height=5cm]{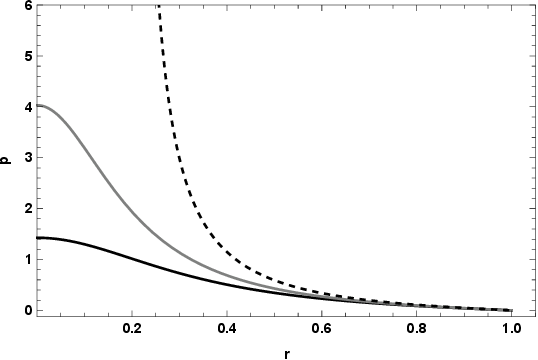}
\caption{Numerical evolution of the internal pressure profile $p(r)$ for the non-interacting case ($\chi \equiv 0$) with a fixed matter density $\rho_0 = 0.10$ and unit radius ($\gamma = 1$). The profiles are shown for increasing values of the vacuum energy density: $V_0 = 0.01$ (black solid line), $V_0 = 0.011$ (grey solid line), and $V_0 = 0.012$ (black dashed line). The divergence of the central pressure for $V_0 = 0.012$ indicates that the configuration has reached the classical Buchdahl limit, serving as our benchmark threshold for the General Relativistic regime. }
\label{fig1}
\end{figure}

With the classical benchmark established, we now shift our focus to the interacting case by introducing a non-zero coupling constant. Specifically, we fix the coupling at $\chi = -0.01$ while maintaining the matter density at $\rho_0 = 0.10$. This specific configuration serves as a primary example to demonstrate how the energy exchange between the matter and vacuum components modifies the hydrostatic equilibrium of the stellar interior. As illustrated in Fig. 2, the introduction of this coupling effectively relaxes the pressure gradient throughout the stellar interior. Most notably, even as the vacuum energy density $V_0$ reaches the previously established threshold of $0.012$ -- where the GR pressure formally diverged -- the central pressure in the interacting case remains finite and well-behaved. It is worth noting that all the parameters used to construct Fig. 2 satisfy (\ref{mbl}) as one should expect.
\begin{figure}[tbp]
\includegraphics[width=8cm,height=5cm]{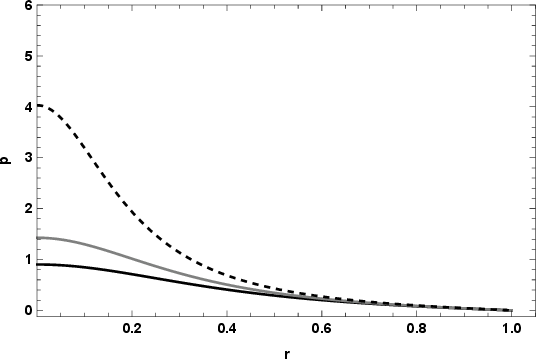}
\caption{Numerical evolution of the internal pressure profile $p(r)$ for the interacting case ($\chi = -0.01$) with a fixed matter density $\rho_0 = 0.10$ and unit radius ($\gamma = 1$). The profiles are illustrated for the same vacuum energy densities as in Fig. 1: $V_0 = 0.01$ (black solid line), $V_0 = 0.011$ (grey solid line), and $V_0 = 0.012$ (black dashed line). In contrast to the non-interacting regime, the profile for $V_0 = 0.012$ remains finite and well-behaved at the origin, demonstrating that the interaction term successfully bypasses the classical Buchdahl limit.}
\label{fig1}
\end{figure}
The comparison between Fig. 1 and Fig. 2 clearly demonstrates that the interaction between the vacuum and matter components alters the stability landscape of the stellar interior. By introducing a coupling of $\chi = -0.01$, the central pressure divergence -- which traditionally marks the non-physicality of a GR configuration -- is successfully suppressed at the classical Buchdahl threshold of $V_0 \simeq 0.012$.
\begin{figure}[tbp]\centering\includegraphics[width=7.5cm,height=5cm]{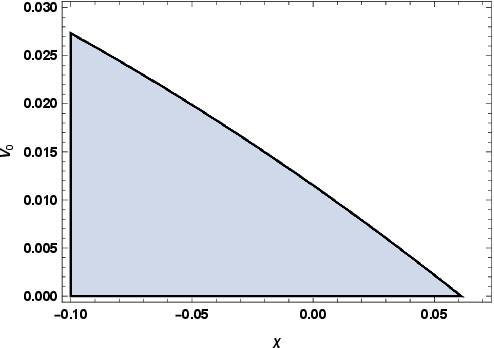}\caption{A diagram illustrating the modified compactness limit 
(\ref{mbl}) considering the radial normalization $\gamma=1$. The shaded portion of the diagram represents the region of allowed stellar configurations, delineating the physical states that satisfy the modified compactness limit (\ref{mbl})}\label{figchi}\end{figure}

In order to provide a more thorough exploration of the parameter space,
in Fig. 3 we show a diagram illustrating the modified compactness limit 
(\ref{mbl}) considering the radial normalization $\gamma=1$. In this diagram one may notice the shift in the stability boundary by comparing the interacting case to the non-interacting General Relativity counterpart, where $\chi=0$ and $V_0 \simeq 0.012$.

This results confirms that a non-vanishing interaction term $Q_\nu$ allows for configurations that would be geometrically forbidden under standard General Relativity, thereby effectively bypassing the classical Buchdahl limit.

\subsection{$Q_\nu={^{(2)}}Q_\nu$}

Having explored the matter-driven interaction, we now turn our attention to the second phenomenological model, where the vacuum interaction is coupled directly to the Ricci scalar. In this case a direct integration of (\ref{eq4}) furnishes
\begin{eqnarray}
\label{solvn}
V(r)=V_0+8\pi\chi\Big[\frac{\rho-3p}{1-32\pi\chi}\Big].    
\end{eqnarray}
where we have used (\ref{eq2}). Therefore
\begin{eqnarray}
\nonumber
{\cal V}(r)=\frac{4}{3}\pi V_0 r^3+\frac{8\chi}{1-32\pi\chi} M(r)\\
-\frac{96\pi^2\chi}{1-32\pi\chi}\int p(\xi) \xi^2 d\xi
\end{eqnarray}
and
\begin{eqnarray}
\nonumber
\bar{v}= \frac{4}{3}\pi V_0+\frac{8\chi}{1-32\pi\chi} \bar{\rho}~~~~~~~~~~~~~~~~~~\\
\label{vbr}
-\frac{96\pi^2\chi}{r^3(1-32\pi\chi)}\int p(\xi) \xi^2 d\xi.    
\end{eqnarray}

At this stage, it is crucial to consider the behavior of the internal fluid under extreme compression. In standard General Relativity, as the compactness of a constant-density star approaches the limit (\ref{bds}), the gradient required to maintain hydrostatic equilibrium forces the central pressure to climb rapidly. This inevitably leads to a regime where the pressure exceeds the energy density ($p > \rho$), signaling a formal violation of the Dominant Energy Condition (DEC). Since we are specifically tracking limiting configurations that may exceed the classical Buchdahl bound, the pressure-integral term in (\ref{vbr}) dominates the interaction. Neglecting the second term on the RHS of (\ref{vbr}) relative to the third, the master equation (\ref{w1}) is modified to:
\begin{eqnarray}
\nonumber
\frac{d}{dr}\Big[\frac{1}{r\sqrt{h}}\frac{d}{dr}(\sqrt{f})\Big]\simeq-\frac{96\pi^2\chi}{(1-32\pi\chi)}\sqrt{fh}\\
\label{w1n}
\times
\frac{d}{dr}\Big[\frac{1}{r^3}\int p(\xi) \xi^2 d\xi\Big].   
\end{eqnarray}    
From (\ref{w1n}) we notice that the existence and value of a compactness limit is no longer an universal geometric property. Instead, the limit becomes dynamical and depends explicitly on the internal physics of the configuration. In standard General Relativity, the Buchdahl limit is Equation of State (EoS) independent. However, in our model, the modified master equation (\ref{w1n})
demonstrates that the geometry is now coupled to the gradient of the integrated pressure. This leads to two primary interpretations depending on the choice of EoS and the interaction 4-vector: 
\\
\\
(i) For a given EoS, the interaction term $\chi$ modifies the hydrostatic equilibrium such that the divergence of the central pressure $p(0)$ is deferred. In this regime, the Buchdahl bound is effectively shifted to a different compactness limit.
\\
\\
(ii) If the interaction term scales sufficiently with the gravitational potential, it can counteract the non-linear growth of the pressure gradient entirely. In such cases, the RHS of Eq. (\ref{w1n}) remains positive-definite and prevents the mathematical singularity in the pressure, potentially allowing for regular configurations.
\\
\\
In conclusion, we argue that the maximum allowable compactness is now a model-dependent bound that emerges from the competition between the fluid's EoS and the coupling parameter $\chi$.

In the following we shall give a numerical example which is connected to the case (ii), that is, assuming the internal pressure -- and consequently its average value -- monotonically decreases toward the stellar surface,
the, the RHS of (\ref{w1n}) becomes a positive-definite function. This result directly violates the classical inequality (\ref{in1}), providing the mathematical mechanism by which the interacting vacuum component bypasses the Buchdahl limit (\ref{bds}).
To evaluate the impact of a curvature-dependent interaction on the stellar stability, we again perform a numerical inward integration of the internal pressure profile, starting from the stellar surface. Our objective is to determine whether the geometric coupling in $^{(2)}Q_\nu$ can yield regular configurations where standard General Relativity predicts a singularity. Following the same protocol established in the previous section, we maintain a matter density of $\rho_0 = 0.10$ and unit radius $\gamma = 1$. This consistency allows for a direct comparison between the results of the previous section and the curvature-driven interaction case. We use the same baseline presented in Fig. 1, where the non-interacting case ($\chi = 0$) reaches the classical Buchdahl limit at a critical vacuum density of $V_0 \simeq 0.012$. Introducing the curvature coupling by fixing $\chi = 0.001$
we provide an example that serves to illustrate how the vacuum, when linked directly to the Ricci scalar, responds to the intense gravitational field near the stability threshold. As shown in Fig. 3, the presence of this interaction term alters the internal pressure distribution. Even as the vacuum energy density reaches the critical threshold of $V_0 = 0.012$, the central pressure remains finite and well-behaved, in stark contrast to the divergence observed in the classical regime illustrated in Fig. 1.
\begin{figure}[tbp]\centering\includegraphics[width=8cm,height=5cm]{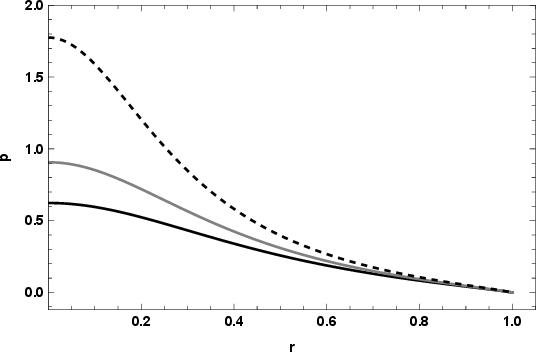}\caption{Numerical evolution of the internal pressure profile $p(r)$ for the curvature-interacting case ($\chi = 0.001$) with fixed $\rho_0 = 0.10$ and $\gamma = 1$. The curves correspond to $V_0 = 0.01$ (black solid), $V_0 = 0.011$ (grey solid), and $V_0 = 0.012$ (black dashed). Unlike the GR case, the curvature coupling prevents the divergence at the origin for $V_0 = 0.012$, effectively extending the stability region.}\label{fig3}\end{figure}

The comparison between the non-interacting benchmark and Fig. 3 demonstrates that the geometric interaction successfully suppresses the central pressure divergence. This result is particularly noteworthy as it confirms that the Buchdahl wall is not an absolute geometric barrier, but rather a limit that can be shifted when the vacuum is allowed to exchange energy with the spacetime curvature. By allowing the vacuum to respond to the Ricci scalar, the system can support ultra-compact configurations that would be otherwise forbidden, providing a clear phenomenological signature of the $^{(2)}Q_\nu$ interaction.

\section{Final Remarks}

In this paper, we have explored the impact of an interacting vacuum component on the hydrostatic equilibrium and stability of ultra-compact stellar configurations. By extending the Tolman-Oppenheimer-Volkoff equations to incorporate a covariant energy exchange term $Q_\nu$, we have demonstrated that the classical Buchdahl limit—traditionally viewed as an absolute geometric barrier in General Relativity -- can be significantly relaxed or bypassed. Our analysis of both matter-driven and curvature-coupled interaction models reveals that the vacuum component does not merely act as a background cosmological constant, but rather as a dynamical agent capable of redistributing gravitational stress and suppressing the divergence of internal pressure.

The numerical results presented in Sections III provide clear phenomenological signatures of these interactions. In the non-interacting regime ($\chi = 0$), the central pressure formally diverges as the system approaches the modified Buchdahl threshold for a given vacuum energy density $V_0$. However, the introduction of a non-zero coupling constant allows for regular, finite pressure profiles even within these previously singular regimes. Specifically, we found that a coupling to matter gradients or to the Ricci scalar provides the necessary restorative effect to maintain equilibrium at high compactness. This suggests that the Buchdahl wall is sensitive to the underlying conservation laws and can be shifted through the inclusion of vacuum-matter coupling, similar to results found in modified gravity frameworks like $f(R)$ or $f(R,T)$ theories \cite{Astashenok:2013vza, Moraes:2015uxq}.

Furthermore, our investigation into the behavior of energy conditions near the stability threshold reveals that interacting models can support configurations that formally violate the Dominant Energy Condition ($p > \rho$) without succumbing to gravitational collapse. In standard General Relativity, such violations are typically precursors to total instability; yet, in our framework, the interaction term effectively softens the gravitational requirement, permitting well-behaved interiors for ultra-compact objects. This provides a compelling theoretical basis for the existence of exotic compact objects, such as gravastars or dark energy stars, which may possess mass-to-radius ratios that defy classical expectations \cite{Mazur:2001fv,Beltracchi:2018ait}.

In conclusion, the interacting vacuum scenario offers a robust framework for investigating new physics in the strong-field regime. By bridging the gap between cosmological dark energy models and relativistic astrophysics \cite{Wang:2016lxa}, we have shown that energy exchange mechanisms play a vital role in determining the fundamental limits of stellar existence. Future work should focus on the stability of these configurations under radial perturbations \cite{Chandrasekhar:1964zza} and the potential gravitational-wave signatures that might arise from such ultra-compact objects. As high-precision data from LIGO-Virgo-KAGRA and NICER continue to refine our understanding of the neutron star equation of state \cite{Miller:2021qha}, the role of vacuum interactions may prove essential in reconciling theoretical stability limits with astrophysical observations.

\section{Acknowledgments}

RM acknowledges financial support from
FAPERJ Grant No. E-$26/010.002481/2019$.

\end{document}